\documentclass[10pt]{report}
\usepackage{graphicx}
\usepackage{rotating}
\usepackage[section]{placeins}

\begin{document}

\Large
\centerline{\bf Evidence of N$_2$-Ice On the Surface of} 

\centerline{\bf the Icy Dwarf Planet 136472 (2005 FY9)} 

\vskip 0.5truein

\normalsize
\centerline{S. C. Tegler$^1$} 
\centerline{Dept. Physics \& Astronomy, Northern Arizona Univ, 
Flagstaff, AZ 86011}
\centerline{E-mail: Stephen.Tegler@nau.edu}
\vskip 0.5truein

\centerline{W. M. Grundy$^1$} 
\centerline{Lowell Observatory, 1400 W. Mars Hill Rd., 
Flagstaff, AZ 86001}
\centerline{E-mail: W.Grundy@lowell.edu}
\vskip 0.5truein

\centerline{F. Vilas$^1$} 
\centerline{MMT Observatory, PO Box 210065, University of Arizona, Tucson, AZ, 85721}
\centerline{Email: fvilas@mmto.org}
\vskip 0.5truein

\centerline{W. Romanishin$^1$}
\centerline{Dept. Physics \& Astronomy, 
Univ of Oklahoma, Norman, OK 73019}
\centerline{E-mail: wjr@nhn.ou.edu}
\vskip 0.5 truein
\vfil\eject

\centerline{D. Cornelison}
\centerline{Dept. Physics \& Astronomy, Northern Arizona Univ, 
Flagstaff, AZ 86011}
\centerline{E-mail: David.Cornelison@nau.edu}
\vskip 0.5truein

\centerline{G. J. Consolmagno$^1$, S.J}
\centerline{Vatican Observatory, Specola, Vaticana, V-00120, Vatican City State}
\centerline{Email: gjc@specola.va}
\vskip 0.7truein

\centerline{Pages: 33; Tables 2; Figures: 7}
\vskip 3.0truein

$^1$ Observer at the MMT Observatory. Observations reported here were
obtained at the MMT Observatory, a joint facility of the University of
Arizona and the Smithsonian Institution.

\vfil\eject

\noindent{Proposed Running Head: \quad N$_2$-Ice On 2005 FY9 \hfil}
\vskip 0.5truein

\noindent{Editorial Correspondence to: \hfil}

\noindent{Dr. Stephen C. Tegler \hfil}

\noindent{Dept Physics \& Astronomy \hfil}

\noindent{Northern Arizona University \hfil}

\noindent{Flagstaff, AZ 86011 \hfil}

\noindent{Phone: 928-523-9382 \hfil}

\noindent{FAX: 928-523-1371 \hfil}

\noindent{E-mail: Stephen.Tegler@nau.edu \hfil}

\vfil\eject

\null

\noindent {\bf Abstract \hfil}
\smallskip

We present high signal precision optical reflectance spectra of 2005
FY9 taken with the Red Channel Spectrograph and the 6.5-m MMT
telescope on 2006 March 4 UT (5000$-$9500 \AA; 6.33 \AA\ pixel$^{-1}$)
and 2007 February 12 UT (6600$-$8500 \AA; 1.93 \AA\
pixel$^{-1}$). From cross correlation experiments between the 2006
March 4 spectrum and a pure CH$_4$-ice Hapke model, we find the
CH$_4$-ice bands in the MMT spectrum are blueshifted by 3 $\pm$ 4 \AA\
relative to bands in the pure CH$_4$-ice Hapke spectrum.  The higher
resolution MMT spectrum of 2007 February 12 UT enabled us to measure
shifts of individual CH$_4$-ice bands. We find the 7296 \AA, 7862 \AA,
and 7993 \AA\ CH$_4$-ice bands are blueshifted by 4 $\pm$ 2 \AA, 4
$\pm$ 4 \AA, and 6 $\pm$ 5 \AA. From four measurements we report here
and one of our previously published measurements, we find the
CH$_4$-ice bands are shifted by 4 $\pm$ 1 \AA.  This small shift is
important because it suggest the presence of another ice component on
the surface of 2005 FY9. Laboratory experiments show that CH$_4$-ice
bands in spectra of CH$_4$ mixed with other ices are blueshifted
relative to bands in spectra of pure CH$_4$-ice. A likely
candidate for the other component is N$_2$-ice because its weak 2.15
$\mu$m band and blueshifted CH$_4$ bands are seen in spectra of Triton
and Pluto. Assuming the shift is due to the presence of N$_2$, spectra
taken on two consecutive nights show no difference in CH$_4$/N$_2$. In
addition, we find no measureable difference in CH$_4$/N$_2$ at
different depths into the surface of 2005 FY9.
\bigskip

\noindent{Key Words: Kuiper Belt Objects, Spectroscopy,
Trans-Neptunian Objects \hfil} 
\vfil\eject

\noindent {\bf 1. Introduction \hfil}
\smallskip

For more than 60 years, Pluto was the only known icy dwarf planet in
the outer Solar System.  The recent discovery of several more of them
opens the door to comparative planetology studies of these small icy
planets.

Pluto, Eris, and (136472) 2005 FY9 form a natural class for
comparative studies, since the near-infrared spectra of all three are
dominated by strong methane ice absorption bands (Owen et al.\ 1993;
Brown et al.\ 2005; Licandro et al.\ 2006a).  The existence of
CH$_4$-ice on their surfaces, despite its relatively short life
expectancy against photolysis and radiolysis, suggests that these
objects experience similar surface and atmospheric activity.  To this
sample of objects with CH$_4$-ice and apparently young surfaces, we
can add Neptune's satellite Triton (likely to have had similar origins
prior to its capture by Neptune) and perhaps Sedna (e.g., Barucci et
al.\ 2005).

In addition to CH$_4$-ice absorptions, the 2.15 $\mu$m band of
nitrogen ice has also been reported on some of these objects (e.g.,
Cruikshank et al.\ 1993; Owen et al.\ 1993; Barucci et al.\ 2005).
Since the N$_2$ absorption is intrinsically much weaker than the
CH$_4$-ice absorptions, its detection requires abundant N$_2$-ice.
N$_2$-ice is far more volatile than CH$_4$-ice.  So where N$_2$-ice
exists, it can be expected to dominate vapor pressure supported
atmospheric compositions relative to CH$_4$.  Much more N$_2$ than
CH$_4$ can be sublimated, transported, and condensed on seasonal time
scales (e.g., Spencer et al.,\ 1997).  That the more volatile N$_2$
ice does not simply bury the less mobile CH$_4$-ice hints at a
complicated interplay between these two ices, which are miscible
within one another (Prokhvatilov and Yantsevich 1983).

Laboratory studies by Quirico and Schmitt (1997) showed that the
wavelengths of near-infrared CH$_4$ ice absorption bands shift to
shorter wavelengths when the CH$_4$ is diluted within N$_2$ ice.  For
ice mixtures with 0.1\% $<$ CH$_4$ $<$ 2\% they reported blueshifts of
$\sim$ 0.006 $\mu$m (60 \AA) relative to bands in pure CH$_4$-ice.
The wavelengths of Triton's CH$_4$ bands, blueshifted by $\sim$ 0.007
$\mu$m (70 \AA), are consistent with its CH$_4$ being highly diluted
in N$_2$ ice (Quirico et al.\ 1999).  Pluto's CH$_4$ wavelength shifts
are smaller, suggesting that Pluto's CH$_4$ occurs in a mixture of
both diluted and undiluted phases (Dout\'e et al.\ 1999; Grundy and
Buie,\ 2001).  Barucci et al. (2005) cross-correlation experiments
between spectra of Sedna and Triton suggest Sedna's CH$_4$-ice bands
have a blueshift similar to Triton's CH$_4$-ice bands.  Licandro et
al.\ (2006b) found blueshifts of 0 \AA\ and 16 \AA\ for the 7296 \AA\
and 8897 \AA\ bands in spectra of Eris. They suggest the different
shifts reflect a decrease in the N$_2$ concentration as a function of
depth into the surface of Eris.

If blueshifts of CH$_4$-ice bands correlate with the amount of
N$_2$-ice, 2005 FY9 may have a surface richer in CH$_4$-ice.  Licandro
et al.\ (2006a) measured blueshifts of only 0 - 6 \AA\ for CH$_4$
bands at optical wavelengths (see their Table 1), but questioned the
reality of these shifts because they were so small. Our group
independently measured blueshifts of 3 $\pm$ 2 \AA\ for the same
bands, thereby bolstering the case for the reality of the small shifts
(Tegler et al.\ 2007). We note a possible typo in the wavelength of
the 3$\nu_1$ $+$ 4$\nu_4$ band for pure CH$_4$-ice in Table 1 of
Licandro et al., we think it should read 7296 \AA\ rather than 7299
\AA, and if so they see no shift for this band.

Recent laboratory measurements confirm a correlation between
blueshifts of CH$_4$ bands and the amount of N$_2$-ice, even at much
lower concentrations of N$_2$ than had previously been investigated.
Cornelison et al.\ (2007) found that a CH$_4$ concentration of 98 \%
(i.e., a N$_2$ concentration of 2\%) exhibited a small, but
measureable, blueshift of 0.0002 $\mu$m (2 \AA) in the $\nu_1$ $+$
$\nu_4$ and $\nu_3$ $+$ $\nu_4$ CH$_4$-ice bands at 2.3789 $\mu$m and
2.3234 $\mu$m.  As the N$_2$ abundance increased, both bands exhibited
progressively larger blueshifts.  For comparison, Quirico and Schmitt
(1997) found that samples with 0.1\% $\le$ CH$_4$ $\le$ 2\% exhibited
blueshifts of 63 \AA, 53 \AA, and 26 \AA\ for the $\nu_1$ $+$ $\nu_4$
band, the $\nu_3$ $+$ $\nu_4$ band, and the $2\nu_1$ $+$ $\nu_3$ $+$
2$\nu_4$ band at 8897 \AA.

Near-infrared spectra of 2005 FY9 are consistent with it having a
CH$_4$-ice rich surface with trace amounts of N$_2$-ice.  Brown et
al.\ (2007) found unusually long optical path lengths through
CH$_4$-ice, $\sim$ 1 cm, compared to 100 $\mu$m for Pluto. In
addition, they found no evidence of the N$_2$-ice band at 2.15
$\mu$m. Finally, Eluszkiewicz et al.\ (2007) used a microphysical and
radiative transfer model to show that it is very unlikely to have long
optical path lengths in both CH$_4$-ice and N$_2$-ice simultaneously
on the same object.

The surfaces of Triton, Pluto, Eris, Sedna, and 2005 FY9 appear to
exhibit a range of N$_2$/CH$_4$ mixing ratios, with 2005 FY9 possibly
being the richest in CH$_4$-ice of the group.  It is not yet clear why
2005 FY9 should be more CH$_4$-rich than the others, but this
characteristic makes it an especially important target for comparative
planetology studies.

Here we report new observations of 2005 FY9 that we combine with our
previously published observation to arrive at a statistically
significant blueshift of the CH$_4$ bands in the spectrum of 2005 FY9.
In addition, we examine individual CH$_4$ bands of 2005 FY9 for
differing shifts as seen in the spectra of Eris, and hence possible
differences in CH$_4$/N$_2$ as a function of depth.  We find no
measureable difference in CH$_4$/N$_2$ as a function of depth.

\bigskip

\noindent {\bf 2. Observations \hfil}
\smallskip

We obtained spectra of 2005 FY9 on 2006 March 4 UT and 2007 February
12 UT with the 6.5-meter MMT telescope on Mt. Hopkins, AZ and the Red
Channel Spectrograph.  On 2006 March 4, we used a 150 g mm$^{-1}$
grating and a 1 $\times$ 180 arc sec slit that provided wavelength
coverage of 5000$-$9500 \AA\ in first order, a dispersion of 6.33 \AA\
pixel$^{-1}$, and a FWHM resolution of 20.0 \AA. On 2007 February 12,
we used a 600 g mm$^{-1}$ grating and a 1 $\times$ 180 arc sec slit
that provided wavelength coverage of 6600$-$8500 \AA\ in first order,
a dispersion of 1.93 \AA\ pixel$^{-1}$, and a FWHM resolution of 4.7
\AA. In Table 1, we present the UT dates, UT times, airmass values,
exposure times, and gratings of our observations. Table 1 includes our
previously published observations of 2005 FY9 on 2006 March 5 UT (Tegler
et al., 2007).

There were high, thin cirrus clouds and the seeing was $\sim$ 0.9 arc
sec on both nights. 2005 FY9 was placed at the center of the slit and
tracked at its rate.

We used the Image Reduction and Analysis Facility (IRAF) and standard
procedures (Massey et al. 1992) to calibrate and extract
one-dimensional spectra from the two-dimensional spectral
images. Specifically, the electronic bias of each image was removed by
subtracting its overscan as well as a bias picture. Pixel to pixel
sensitivity variations were removed from each image by dividing by a
normalized twilight flatfield image. Extraction of one-dimensional
spectra from the images was done with the apall task in IRAF.  HeNeAr
spectra taken before and after each set of object spectra were used to
minimize the effects of flexure and obtain an accurate wavelength
calibration. Measurements of airglow lines in the spectra indicate our
wavelength scales are accurate to $\sim$ one-tenth of a pixel or
$\sim$ 0.6 \AA\ for the 2006 March 4 observations and 0.2 \AA\ for the
2007 February 12 observations.  We removed telluric bands and
Fraunhofer lines from the 2005 FY9 spectra by observing the solar
analog star HD 112257 (Hardorp, 1982) at airmasses very close to 2005
FY9, and then dividing the 2005 FY9 spectra by the normalized solar
analog spectra. Typically, the airmass difference between 2005 FY9 and
the solar analog star was $\le$ 0.05.

\bigskip

\noindent {\bf 3. Results \hfil} 
\medskip

\noindent {\bf 3.1 Lower Resolution Spectrum of 2006 March 4 UT}
\medskip

First, we compare our spectrum of 2006 March 4 UT to our previously
published spectrum of 2006 March 5 UT.  In Figure 1, we plot the March
4 UT spectrum, a median of six 15-minute spectra corresponding to a
total exposure time of 90 minutes (black line), and the published
March 5 UT spectrum (Tegler et al., 2007), a median of four 10-minute
exposures corresponding to a total exposure time of 40 minutes (red
line).  We combined the images by a median because the new deep
depletion CCD is sensitive to cosmic rays. We assume an albedo of 0.8
at 6500 \AA\ for both spectra. Such a value is consistent with the
albedo of 2005 FY9 measured with the Spitzer Space Telescope
(Stansberry et al. 2007). We see no significant difference between the
two spectra other than the March 4 spectrum has better signal
precision than the March 5 spectrum.  There are no period of rotation
or obliquity measurements for 2005 FY9, and so the difference in
rotational phase between the two nights is unknown. Furthermore, it is
possible that 2005 FY9 has a nearly pole-on aspect.

Next, we compare our spectrum of 2006 March 4 UT to a spectrum of pure
CH$_4$-ice.  Such a comparison is important for three reasons.  First,
it demonstrates that the absorption bands in the spectrum of 2005 FY9
are due to CH$_4$-ice. Second, it demonstrates that the signal
precision of our MMT spectrum rivals the signal precision of laboratory
data. Finally, the comparison makes it possible for us to perform
cross correlation experiments, and thereby measure the blueshifts of
CH$_4$-ice bands in spectra of 2005 FY9.

We used Hapke theory (Hapke 1981, Hapke 1993) to transform laboratory
optical constants of pure CH$_4$-ice at 30 K (Grundy et al. 2002) into
an albedo spectrum for comparison to the MMT spectra.  Specifically,
we fit the MMT spectra of 2006 March 4 and March 5 UT with Hapke
parameters of h $= $0.1, B$_o$ $=$ 0.8, $\overline{\theta}$ $=$
30$^{\circ}$, P(g) $=$ a two component Henyey-Greenstein function with
80\% in the forward scattering lobe and 20 \% in the back scattering
lobe and both lobes having asymmetry parameter 0.63, and two grain
sizes, 6 cm (97\% by volume) and 1 mm (3\% by volume).  These
parameters are comparable to parameters in Hapke model fits of Pluto,
Triton, and other outer Solar System object spectra (e.g. see Grundy and
Buie, 2001). We point out that the parameters do not represent a
unique fit to the MMT spectra; however, they are plausible values for
transparent, pure CH$_4$-ice grains. In Figure 2, we plot our 2006
March 4 UT spectrum (black line) and pure CH$_4$-ice Hapke model
spectrum (red line).  The comparison between this Hapke model and the
published 2006 March 5 UT spectrum is given in Tegler et
al. (2007). We emphasize that our goal here is to use the pure CH$_4$
Hapke model spectrum as a fiducial for blueshift measurements, not to
arrive at unique values for Hapke parameters, and therefore say we
know something definitive about the surface texture of 2005 FY9.

A visual inspection of the two spectra in Figure 2 clearly shows the
absorption bands in the spectrum of 2005 FY9 are due to CH$_4$-ice.
Furthermore, it appears the CH$_4$-ice bands in the MMT spectrum are
slightly blueshifted relative to the bands of the pure CH$_4$-ice
Hapke spectrum.

In order to quantify the apparent shift, we performed a
cross-correlation experiment on the two spectra in Figure 2.  In
particular, we shifted the model spectrum from -12.66 \AA\ to $+$
18.99 \AA\ in 6.33 \AA\ (i.e. 1 pixel) steps. For each shift, we
calculated

\begin{equation}
\chi^2 = {1 \over N}\sum_{i}^{N} {(y_{d,i} - y_{m,i})^2 \over y_{m,i}}
\end{equation}

\noindent where $y_{d,i}$ and $y_{m,i}$ represent the number of
photons in wavelength bin i of the 2005 FY9 and Hapke
spectra. In Figure 3, we present a plot of $\chi^2$ vs. shift. The red
curve is the best fit parabola to the calculated points.  We find the
parabola has a minimum at a shift of 3 \AA.

What is the uncertainty in the 3 \AA\ measurement?  We note that the
HeNeAr spectra enabled us to calibrate the wavelengths in the 2005 FY9
spectrum to an uncertainty of $\sim$ 1/10 of a pixel or $\sim$ 0.6
\AA. A larger source of uncertainty comes from the noise in the
spectrum and the broadness of the cross correlation parabola in Figure
3.  We applied the minimum $\chi^2$ method to arrive at the 1 $\sigma$
confidence region (Avni, 1976). In short, we find the CH$_4$-ice bands
in our 2006 March 4 spectrum of 2005 FY9 are blueshifted relative to
the pure CH$_4$ Hapke model spectrum by 3 $\pm$ 4 \AA, a result
consistent with the blueshift in our 2006 March 5 UT spectrum, 3 $\pm$
2 \AA\ (Tegler et al., 2007).

\medskip

\noindent {\bf 3.2 Higher Resolution Spectrum of 2007 February 12 UT}
\medskip

Next, we compare our higher resolution MMT spectrum of 2007 February
12 UT to a pure CH$_4$-ice Hapke model.  In Figure 4, we plot the MMT
spectrum, a median of ten 10-minute spectra corresponding to a total
exposure time of 100 minutes (black line), and a pure CH$_4$-ice Hapke
model that is nearly identical to the Hapke model in Figure 2 (red
line). In particular, the model here again uses laboratory optical
constants for pure CH$_4$-ice at 30 K (Grundy et al., 2002), Hapke
parameters h $=$ 0.1, B$_o$ $=$ 0.8, $\overline{\theta}$ $=$
30$^{\circ}$, and a two component Henyey-Greenstein function with 80\%
in the forward scattering lobe and 20\% in the back scattering lobe
and both lobes having asymmetry parameter 0.63. We again use grain
sizes of 6 cm and 1 mm; however, here we use slightly different
percentages by volume for the two grain sizes. Specifically, we use
98.3\% and 1.7\% instead of 97\% and 3\%. The major difference between
the two models is the model here gives slightly larger band depths
than the model in section 3.1. In the top, middle, and bottom panels
of Figure 4, we plot the 7296 \AA, 7862 \AA, and 7993 \AA\ CH$_4$-ice
bands. Considering there is no reason to expect the Hapke parameters
to be constants in time or over the surface of 2005 FY9, it is
remarkable that spectra of 2005 FY9 taken nearly a year apart can be
fit by Hapke models with nearly identical parameters.

Again, visual inspection of Figure 4 suggests the CH$_4$ bands in the
MMT spectrum are blueshifted relative to the pure CH$_4$ Hapke
model. Because of the higher spectral resolution, this time we
performed cross correlation experiments on individual bands. We plot
the results of the cross correlation experiments in Figure 5.  After
applying the minimum $\chi^2$ method, we found the 7296 \AA, 7862 \AA,
and 7993 \AA\ CH$_4$ bands in the MMT spectrum are blueshifted by 4
$\pm$ 2 \AA, 4 $\pm$ 4 \AA, and 6 $\pm$ 5 \AA\ relative to the pure
CH$_4$ Hapke model.

In order to test the sensitivity of our cross-correlation results to
small changes in Hapke parameters (specifically the two different models in
Figures 2 and 4), we performed a cross-correlation experiment between
the two model spectra over the 7002 \AA\ to 8505 \AA\ wavelength
range. In Figure 6, we plot $\chi^2$ vs. shift. We find a minimum at
0 $\pm$ 4 \AA, i.e. we find no measureable shift between the two pure
CH$_4$ model spectra.  It appears the slight difference in the model
parameters has no systematic effect on the shifts.

In Table 2, we present a summary of our cross-correlation
measurements. Five independent measurements all indicate the CH$_4$
bands in the spectra of 2005 FY9 are blueshifted relative to CH$_4$
bands in spectra of pure CH$_4$.  The average and standard deviation
of the five measurements in Table 1 are 4 $\pm$ 1 \AA, i.e. we find a
4 $\sigma$ detection of a shift. We interpret the shift as evidence of
trace amounts of N$_2$-ice on the surface of 2005 FY9.

\noindent {\bf 4. Discussion \hfil}
\smallskip  

In this section, we describe how our observations constrain the
CH$_4$/N$_2$ abundance as a function of depth into the surface of 2005
FY9. Then, we compare our results to Eris (Licandro et al., 2006b).
Finally, we put forth a possible mechanism to explain the differences
between 2005 FY9 and Eris.

The average penetration depth of a photon at some wavelength depends
on the reciprocal of the absorption coefficient, i.e.\ larger
absorption coefficients absorb more light, preventing it from
penetrating as deeply into the surface.  Penetration depth also
depends on scattering, which is a complex function of particle sizes,
shapes, and spacing in a particular material, or of void shapes,
sizes, and spacing in a compacted slab as envisioned by Eluszkiewicz
et al. (2007). Photons which go on to be scattered out of a surface
(and potentially be observed) sample shallower depths on average than
the mean penetration depths, reducing the dependence of mean depth
sampled on absorption coefficient.  Nevertheless, weaker absorption
bands do sample, on average, deeper surface strata.

Radiative transfer models can be used to get an idea of the depths
sampled by the spectral observations. As described previously (Tegler
et al., 2007), a Hapke model (Hapke 1993) with a bimodal particle size
distribution can match the observed spectrum of 2005 FY9 reasonably
well. We used 6~cm and 1~mm CH$_4$ ice particles, with mixing ratio
98.3\% to 1.7\% by volume. To match the spectral continuum, an
artificial tholin-like absorber was dispersed within the CH$_4$ ice
particles. These model parameters should not be taken as a unique
solution for the surface texture of 2005~FY9, but rather as plausible
values, consistent with the observed spectra, for purposes of
investigating how deeply within the surface various wavelengths
probe. These model parameters were used in a multiple scattering Monte
Carlo ray tracing model (Grundy and Stansberry, 2000) to explore the
trajectories followed by observable photons of different
wavelengths. The average depths sampled are shown as a function of
wavelength in Figure 7. The depths probed are highly sensitive to the
assumptions made about void space (we assumed a 50\% filling factor,
but this value is merely conjectural $-$ Eluszkiewicz et al's
compacted slab would have a filling factor closer to 100\%) as well as
particle (or void) shape, but the general results that the 7862~\AA\
and 7993~\AA\ absorption bands probe some 30 to 60\% deeper than the
7296~\AA\ band, and a factor of 2 to 3 times deeper than the much
stronger 8897~\AA\ band appear to be relatively robust.

Since we find no measurable difference in the blueshifts of the
7296~\AA, 7862~\AA, and 7993~\AA\ bands of 2005~FY9 (see Figures 4 and
5), we find no measurable difference in the CH$_4$/N$_2$ ice abundance
at the average depths probed by these bands (see Figure 7).

It appears that there is an important difference between the surfaces
of 2005~FY9 and Eris. Licandro et al.\ (2006b) found the 7296~\AA\ and
8897~\AA\ bands of Eris were blueshifted by 0~$\pm$~3~\AA\ and
15~$\pm$~3~\AA, respectively, suggesting an increase in the
CH$_4$/N$_2$ abundance with depth below the surface of Eris. For
2005~FY9, we find no difference in the blueshifts of the 7296~\AA,
7862~\AA, and 7993~\AA\ bands, suggesting a more uniform CH$_4$/N$_2$
abundance with depth.  We note that if the 8897~\AA\ band was
blueshifted by 15~\AA, we would have detected it in our low resolution
spectra of 2005~FY9\null.  Unfortunately, we were clouded out on the
second night of our higher spectral resolution (2007 February) run,
and therefore could not measure a separate blueshift for the 8897~\AA\
band. On the other hand, the similar shifts of the 7296 \AA\ bands of
Eris and 2005 FY9 suggest similar CH$_4$/N$_2$ abundances below the
depth sampled by the 8897 \AA\ band.

What could account for the apparent depth-dependence of CH$_4$/N$_2$
abundance on Eris and absence thereof on 2005~FY9? Both bodies are
currently near aphelion, so the surfaces of both might be expected to
be characteristic of similar phases of their seasonal cycles, with the
most volatile species having condensed last as both objects moved away
from perihelion, as proposed for Eris by Licandro et al.\ (2006b). The
obliquities of Eris and 2005~FY9 are not yet known.  One possible
difference could arise from their orientations. An object oriented
equator-on to the Sun at aphelion would also have had its low
latitudes exposed to sunlight at perihelion, when seasonal volatile
transport would have been most active. Sublimation coupled with solar
gardening on such a body, as envisioned by Grundy and Stansberry
(2000) could account for increasing concentration of CH$_4$ with
depth, as is observed on Eris. Since solar illumination is absorbed,
on average, at greater depths within a surface composed of CH$_4$ and
N$_2$ ices than the depths from which thermal emission escapes,
sublimation would be expected to take place from the
subsurface. CH$_4$ is less volatile than N$_2$, so its concentration
would tend to increase where sublimation acts.  In contrast, if
2005~FY9 is currently nearly pole-on to the Sun at aphelion, the
opposite pole would have been oriented toward the Sun at perihelion,
and the pole we see now would have been the winter hemisphere at
perihelion, making it a depositional environment rather than the
erosional environment which would be expected for the equator-on
orientation.

\vfil\eject

\noindent {\bf Acknowledgements \hfil} 
\bigskip

S.C.T., W.R., and D.C. gratefully acknowledge support from NASA
Planetary Astronomy grant NNG06G138G to Northern Arizona University
and the University of Oklahoma. W.M.G. gratefully acknowledges support
from Planetary Geology and Geophysics grant NNG04G172G to Lowell
Observatory.  We thank Steward Observatory for consistent allocation
of telescope time on the MMT.
\bigskip

\vfil\eject

\noindent{\bf References \hfil}

\noindent Avni, Y., 1976. Energy spectra of x-ray clusters of
galaxies. Astrophys. J. 210, 642-646.
\bigskip

\noindent Barucci, M. A., Cruikshank, D. P., Dotto, E., Merlin, F.,
Poulet, F., Dalle Ore, C., Fornasier, S., and de Bergh, C., 2005. Is
Sedna another Triton? Astron. Astrophys. 439, L1-L4.
\bigskip

\noindent Brown, M. E., Barkume, K. M., Blake, G. A., Schaller, E. L.,
Rabinowitz, D. L., Roe, H. G., and Trujillo, C. A., 2007. Methane and
ethane on the bright Kuiper belt object 2005 FY9. Astron. J. 133,
284-289.
\bigskip

\noindent Brown, M.E., Trujillo, C. A., and Rabinowitz D. L., 2005.  Discovery
of a planetary-sized object in the scattered Kuiper belt.  Astrophys.\
J. 635, L97-L100.
\bigskip

\noindent Cornelison, D., Tegler, S. C., Grundy, W., Crisp, A.,
Abernathy, M., 2007.  Near-infrared spectroscopy of CH$_4$ and N$_2$
ice mixtures: implications for icy dwarf planets. Icarus, submitted.
\bigskip

\noindent Cruikshank, D. P., Roush, T. L., Owen, T. C., Geballe,
T. R., de Bergh, C., Schmitt, B., Brown, R. H., Bartholomew, M. J.,
1993. Ices on the surface of Triton. Science 261, 742-745.
\bigskip

\noindent Dout\'e, S., Schmitt, B., Quirico, E., Owen, T.C., Cruikshank, D. P.,
de~Bergh, C., Geballe, T.R., and Roush, T. L., 1999.  Evidence for
methane segregation at the surface of Pluto.  Icarus 142, 421-444.
\bigskip

\noindent Eluszkiewicz, J., Cady-Pereira, K., Brown, M. E.,
Stansberry, J. A., 2007. Interpretation of the near-ir spectra of the
Kuiper belt object (136472) 2005 FY9. J. Geophys. Res. 112, E06003.
\bigskip

\noindent Grundy, W. M., and Buie, M. W., 2001. Distribution and evolution
of CH$_4$, N$_2$, and CO ices on Pluto's surface: 1995 to 1998. Icarus
153, 248-263.
\bigskip

\noindent Grundy, W.M., and J.A. Stansberry 2000. Solar gardening and
the seasonal evolution of nitrogen ice on Triton and Pluto. Icarus
148, 340-346.
\bigskip

\noindent Grundy, W. M., Schmitt, B., Quirico, E., 2002. The
temperature-dependent spectrum of methane ice I between 0.7 and 5
$\mu$m and opportunities for near-infrared remote thermometry. Icarus
155, 486-496.
\bigskip

\noindent Hapke, B., 1981. Bidirectional reflectance
spectroscopy. 1. Theory. J. Geophys. Res. 86, 4571-4586.
\bigskip

\noindent Hapke, B., 1993. Theory of Reflectance and Emittance
Spectroscopy.  Cambridge Univ. Press, New York.
\bigskip

\noindent Hardorp, J., 1982. The sun among the stars. V - A second
search for solar spectral analogs: The Hyades'
distance. Astron. Astrophys. 105, 120-132.
\bigskip

\noindent Licandro, J., Pinilla-Alonso, N., Pedani, M., Oliva, E.,
Tozzi, G. P., Grundy, W. M., 2006a. The methane ice rich surface of
large TNO 2005 FY9: a Pluto-twin in the trans-neptunian belt?
Astron. Astrophys. 445, L35-L38.
\bigskip

\noindent Licandro, J., Grundy, W. M., Pinilla-Alonso, N., Leisy, P.,
2006b. Visible spectroscopy of 2003 UB313: evidence for N$_2$ ice on
the surface of the largest TNO? Astron. Astrophys. 458, L5-L8.
\bigskip

\noindent Massey, P., Valdes, F., Barnes, J., 1992. A user's guide to
reducing slit spectra with IRAF (Tucson: NOAO),
http://iraf.noao.edu/iraf/ftp/iraf/docs/spect.ps.Z
\smallskip

\noindent Owen, T. C., Roush, T. L., Cruikshank, D. P., Elliot, J. L.,
Young, L. A., de Bergh, C., Schmitt, B., Geballe, T. R., Brown, R. H.,
Bartholomew, M. J., 1993. Surface ices and the atmospheric composition
of Pluto. Science 193, 745-748.
\bigskip

\noindent Prokhvatilov, A. I., and Yantsevich, L. D., 1983.  X-ray
investigation of the equilibrium phase diagram of CH$_4$-N$_2$ solid
mixtures.  Sov.\ J. Low Temp.\ Phys.\ 9, 94-98.
\bigskip

\noindent Quirico, E., and Schmitt, B., 1997. Near-infrared
spectroscopy of simple hydrocarbons and carbon oxides diluted in solid
N$_2$ and as pure ices: implications for Triton and Pluto. Icarus 127,
354-378.
\bigskip

\noindent Quirico, E., Dout\'e, S., Schmitt, B., de~Bergh, C.,
Cruikshank, D. P., Owen, T. C., Geballe, T. R., and Roush, T. L.,
1999.  Composition, physical state, and distribution of ices at the
surface of Triton.  Icarus 139, 159-178.
\bigskip

\noindent Spencer, J. R., Stansberry, J. A., Trafton, L. M., Young,
E. F., Binzel, R. P., and Croft, S. K., 1997.  Volatile transport,
seasonal cycles, and atmospheric dynamics on Pluto.  In {\it Pluto and
Charon} (S. A. Stern and D. J. Tholen, Eds), University of Arizona
Press, Tucson, pp. 435-473.
\bigskip

\noindent Stansberry, J., Grundy, W., Brown, M., Cruikshank, D.,
Spencer, J., Trilling, D., and Margot, J., 2008. Physical properties
of Kuiper belt and Centaur objects: constraints from Spitzer Space
Telescope. In {\it The Solar System Beyond Neptune} (A. Barucci,
H. Boehnhardt, D. Cruikshank, and A. Morbidelli, Eds.), Univ Arizona
Press, Tucson, in press.
\bigskip

\noindent Tegler, S. C., Grundy, W. M., Romanishin, W., Consolmagno,
G. J., Mogren, K., Vilas, F., 2007. Optical spectroscopy of the large
Kuiper belt objects 136472 (2005 FY9) and 136108 (2003
EL61). Astron. J. 133, 526-530.
\bigskip

\vfil\eject

\begin{center}
\noindent\begin{tabular}{ccccc}
\multicolumn{5}{c}{\bf Table 1 } \\ \multicolumn{5}{c}{\bf Dates, Times, Gratings of  2005 FY9 Observations}\\
\hline
\\
UT Date & UT Time$^a$ & Airmass$^a$  & Exp Time & Grating \\
        &  (hh:mm)    &              & (min)    & (g/mm) \\
\\
\hline 
\\ 
2006 Mar 04 & 06:56 & 1.11 & 900 & 150 \\
            & 07:20 & 1.07 & 900 & 150 \\
	    & 07:47 & 1.04 & 900 & 150 \\
            & 08:03 & 1.02 & 900 & 150 \\
            & 08:28 & 1.01 & 900 & 150 \\
            & 08:44 & 1.00 & 900 & 150 \\
2006 Mar 05$^b$ & 05:08 & 1.49 & 600 & 150 \\
            & 05:31 & 1.37 & 600 & 150 \\
            & 06:05 & 1.24 & 600 & 150 \\
            & 06:49 & 1.12 & 600 & 150 \\
2007 Feb 12 & 10:01 & 1.00 & 600 & 600 \\
            & 10:11 & 1.00 & 600 & 600 \\
            & 10:27 & 1.00 & 600 & 600 \\
            & 10:38 & 1.00 & 600 & 600 \\
            & 11:19 & 1.02 & 600 & 600 \\
            & 11:29 & 1.03 & 600 & 600 \\
            & 11:39 & 1.05 & 600 & 600 \\
            & 12:22 & 1.11 & 600 & 600 \\
            & 12:32 & 1.13 & 600 & 600 \\
            & 12:43 & 1.16 & 600 & 600 \\
\\
\hline 
\end{tabular}
\end{center}
\indent {$^a$ Values at beginning of exposures. \hfil} \\
 \qquad \qquad \qquad {$^b$ Tegler et al. (2007). \hfil \\

\vfil\eject

\begin{center}
\noindent\begin{tabular}{lcc}
\multicolumn{3}{c}{\bf Table 2 } \\ \multicolumn{3}{c}{\bf Blueshifts of CH$_4$-Ice Bands}\\
\hline
\\
UT Date & Cross-Corr Range & Blueshift \\
        &  (\AA)   &  (\AA)\\
\\
\hline 
\\ 
2006 Mar 04 & 7020$-$9280 & 3 $\pm$ 4\\
2006 Mar 05$^a$ & 7020$-$9280 & 3 $\pm$ 2\\
2007 Feb 12 & 7150$-$7400 & 4 $\pm$ 2\\
            & 7800$-$7900 & 4 $\pm$ 4\\
            & 7920$-$8020 & 6 $\pm$ 5\\
 
\\
\hline 
\end{tabular}
\end{center}
\indent \qquad \qquad \qquad {$^a$ Tegler et al. (2007).\hfil} \\

\vfil\eject

\noindent {\bf Figure Captions \hfil}

\bigskip

\noindent {\bf Fig. 1.} Portions of our lower resolution albedo
spectra of 2005 FY9 taken with the 6.5-m MMT telescope and the Red
Channel Spectrograph on 2006 March 4 UT (black line) and 2006 March 5
UT (red line). Tick marks indicate maximum absorptions of pure
CH$_4$-ice bands at 7296, 7862, 7993, 8415, 8442, 8691, 8897, 8968,
and 9019 \AA. We find no measureable difference between the CH$_4$-ice
bands on two consecutive nights.
\bigskip

\noindent {\bf Fig. 2.} Our MMT 2006 March 4 UT albedo spectrum of
2005 FY9 (black line) and a pure CH$_4$-ice Hapke model with grain
sizes of 6 cm and 1 mm (red line). Tick marks indicate maximum
absorptions of pure CH$_4$-ice bands. A visual inspection suggests the
CH$_4$ bands in the MMT spectrum are slightly blueshifted relative to
the bands in the pure CH$_4$-ice Hapke model.
\bigskip

\noindent {\bf Fig. 3.} Results of cross-correlation experiments
between the 2006 March 4 MMT spectrum of 2005 FY9 and the pure
CH$_4$-ice Hapke model in Fig. 2. The $\chi^2$ parabola reaches a
minimum at a 3 $\pm$ 4 \AA\ blueshift of the data relative to the
model.
\bigskip

\noindent {\bf Fig. 4.} Three portions of our higher resolution 2007
February 12 MMT spectrum of 2005 FY9 (black line) and a pure
CH$_4$-ice Hapke spectrum with grain sizes of 6 cm and 1 mm (red
line). Top, middle, and bottom panels show the 7296 \AA, 7862 \AA, and
7993 \AA\ CH$_4$-ice bands.  Tick marks indicate the wavelengths of
maximum absorption by pure CH$_4$-ice. A visual inspection suggests
the three CH$_4$-ice bands in the MMT spectrum are slightly
blueshifted relative to the bands in the pure CH$_4$-ice Hapke model.
\bigskip

\noindent {\bf Fig. 5.} Results of cross-correlation experiments
between our 2007 February 12 MMT and pure CH$_4$-ice Hapke spectra in
Figure 4.  The $\chi^2$ parabolas for the 7296 \AA\ (top panel), 7862
\AA\ (middle panel), and 7993 \AA\ (bottom panel) bands reach minima
at 4 $\pm$ 2 \AA, 4 $\pm$ 4 \AA, and 6 $\pm$ 5 \AA\ blueshifts of the
data relative to the model.
\bigskip

\noindent {\bf Fig. 6.} Results of cross-correlation experiments
between two nearly identical pure CH$_4$-ice Hapke spectra in Figures
2 and 4. The parabola has a minimum at 0 $\pm$ 4 \AA. 
\bigskip

\noindent {\bf Fig. 7.} Mean depth sampled by observable photons from
2005~FY9 as a function of wavelength, from a Monte Carlo ray-tracing
model (red line).  The depth sampled depends a number of poorly
constrained factors, such as the porosity and the shapes of the
particles. Higher porosity, or more forward scattering particles
increase the mean depth sampled.  The effect of a $\pm$5\% albedo
uncertainty is indicated by the width of the gray zone. Considering
how little is known at present, the depths here are representative,
rather than definitive. The important things to note from this figure
are the relative depths probed by various CH$_4$ ice absorption bands.

\vfil\eject

\begin{figure}
        \centering
        \scalebox{0.7}{\includegraphics{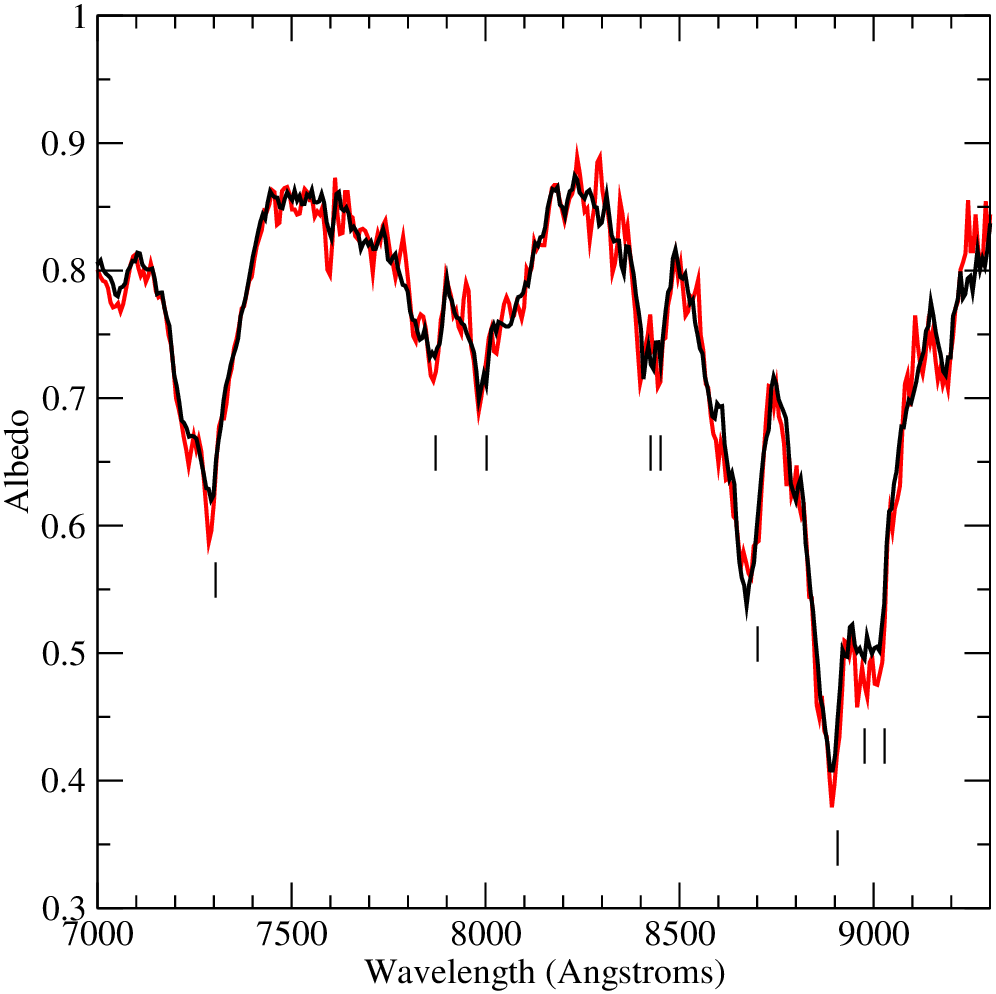}}
        \end{figure}

\begin{figure}
	\centering
	\scalebox{0.7}{\includegraphics{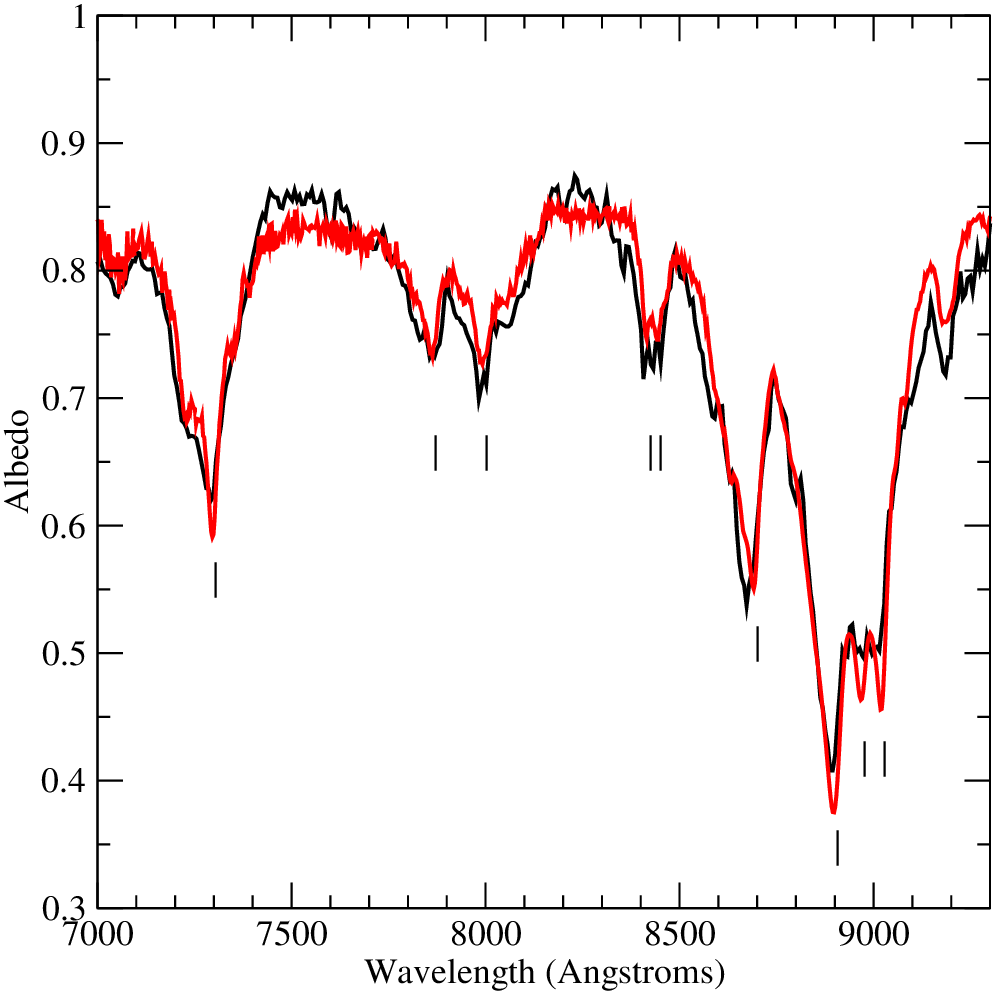}}
	\end{figure}

\begin{figure}
	\centering
	\scalebox{0.7}{\includegraphics{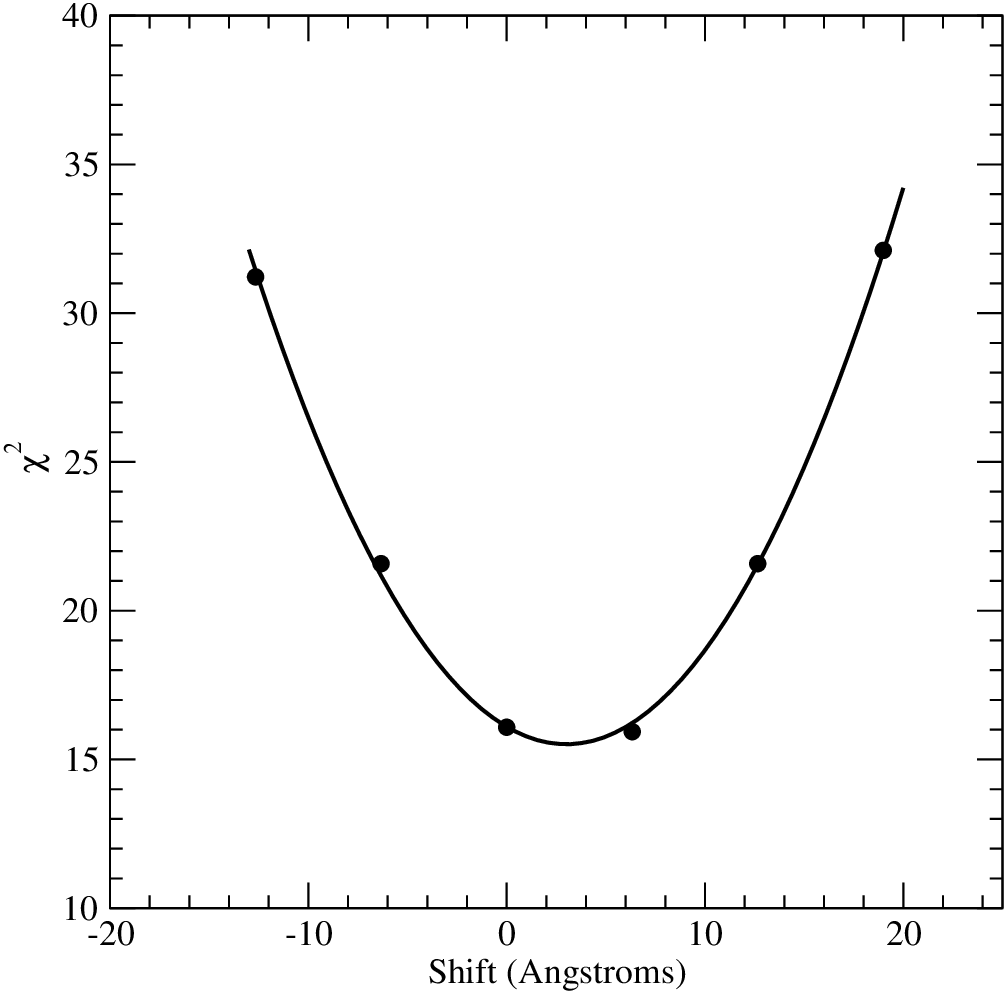}}
	\end{figure}

\begin{figure}
	\centering
	\scalebox{0.7}{\includegraphics{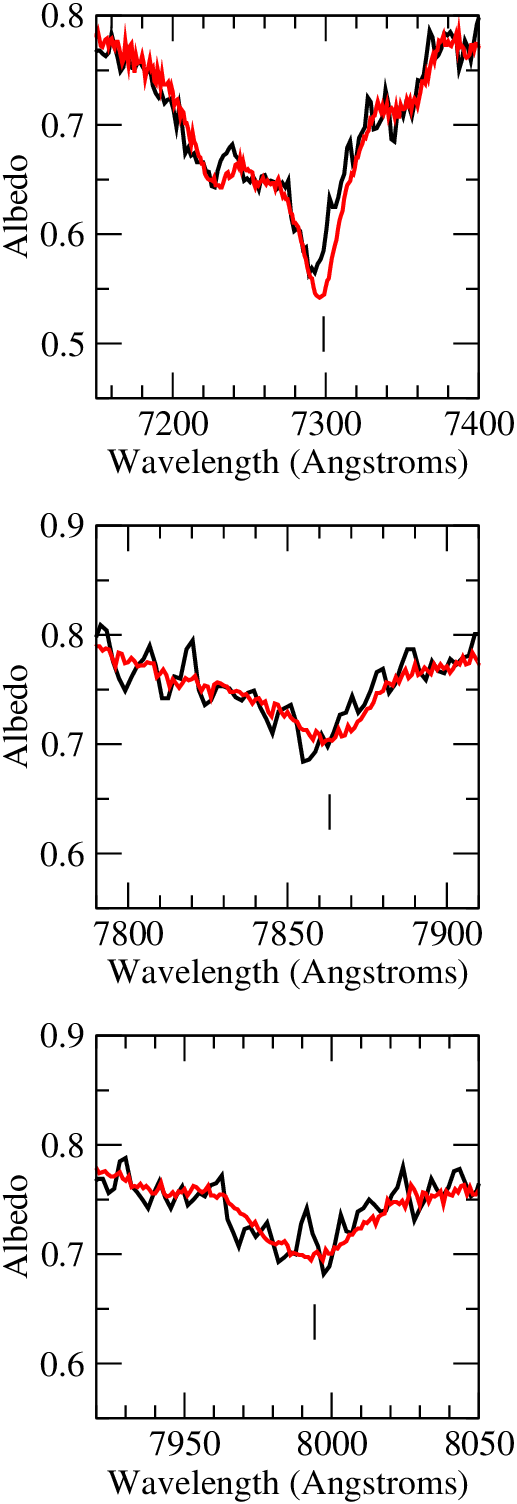}}
	\end{figure}

\begin{figure}
	\centering
	\scalebox{0.7}{\includegraphics{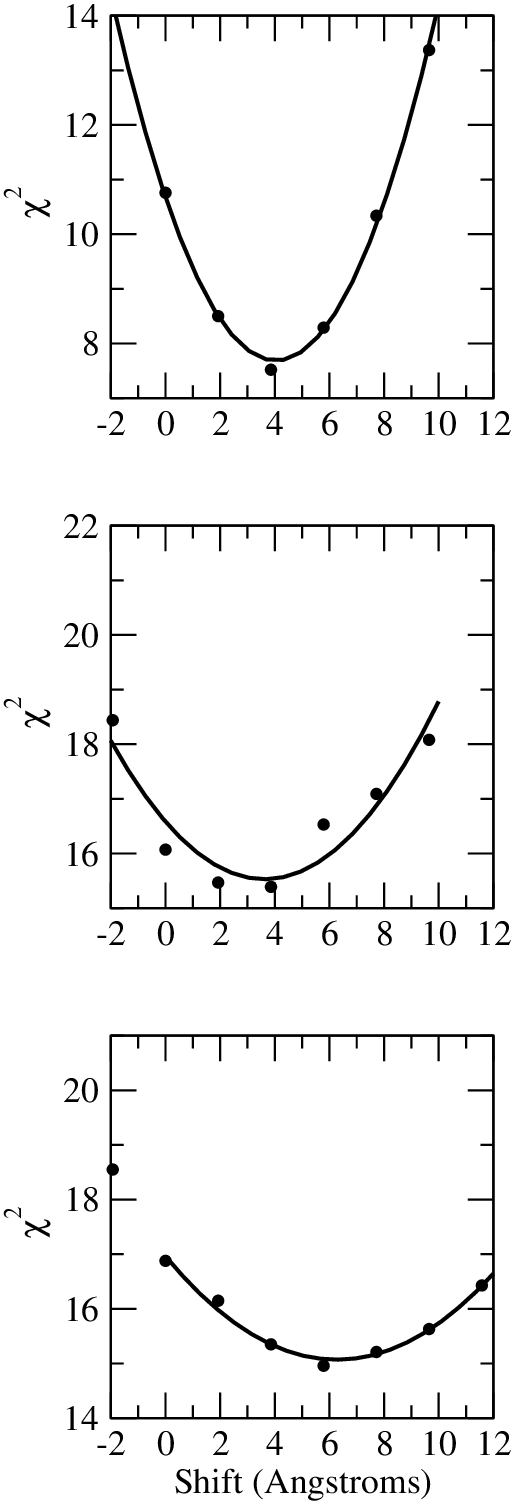}}
	\end{figure}

\begin{figure}
	\centering
	\scalebox{0.7}{\includegraphics{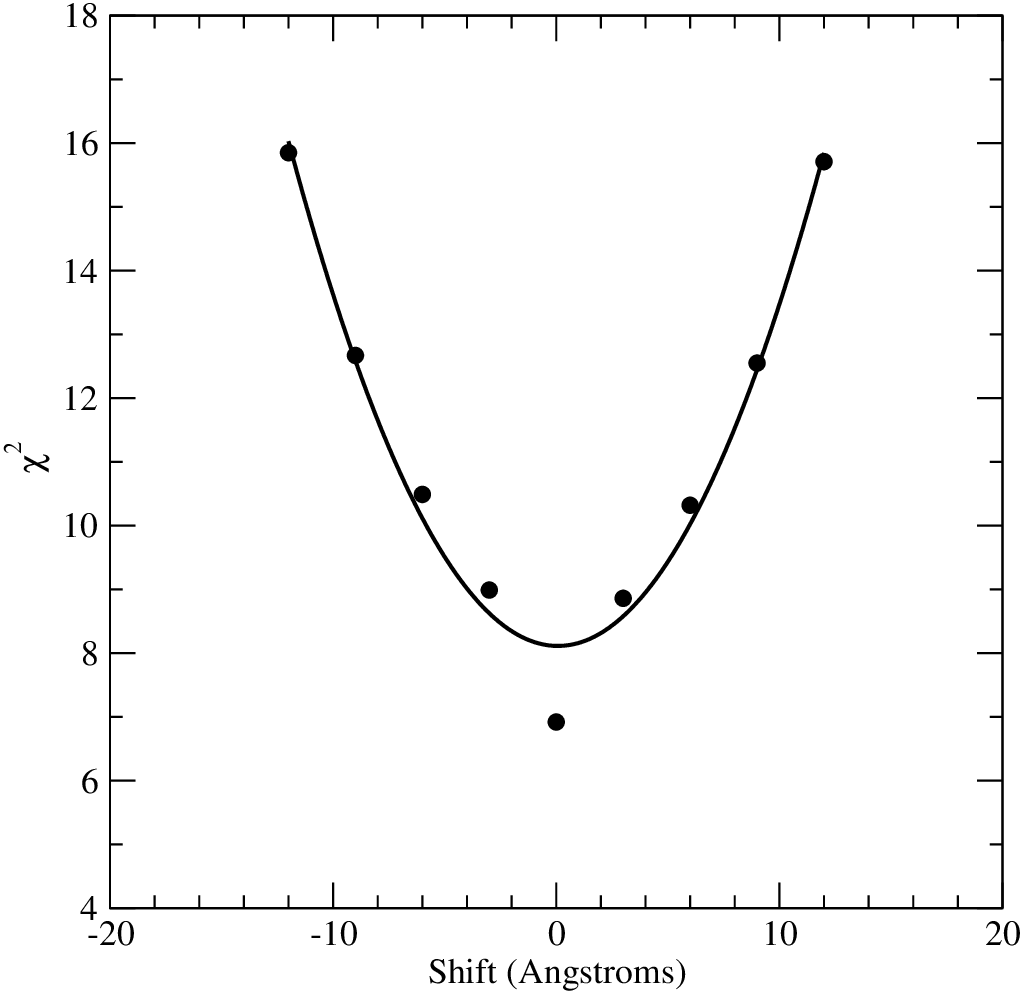}}
	\end{figure}

\begin{figure}
	\centering
	\scalebox{0.7}{\includegraphics{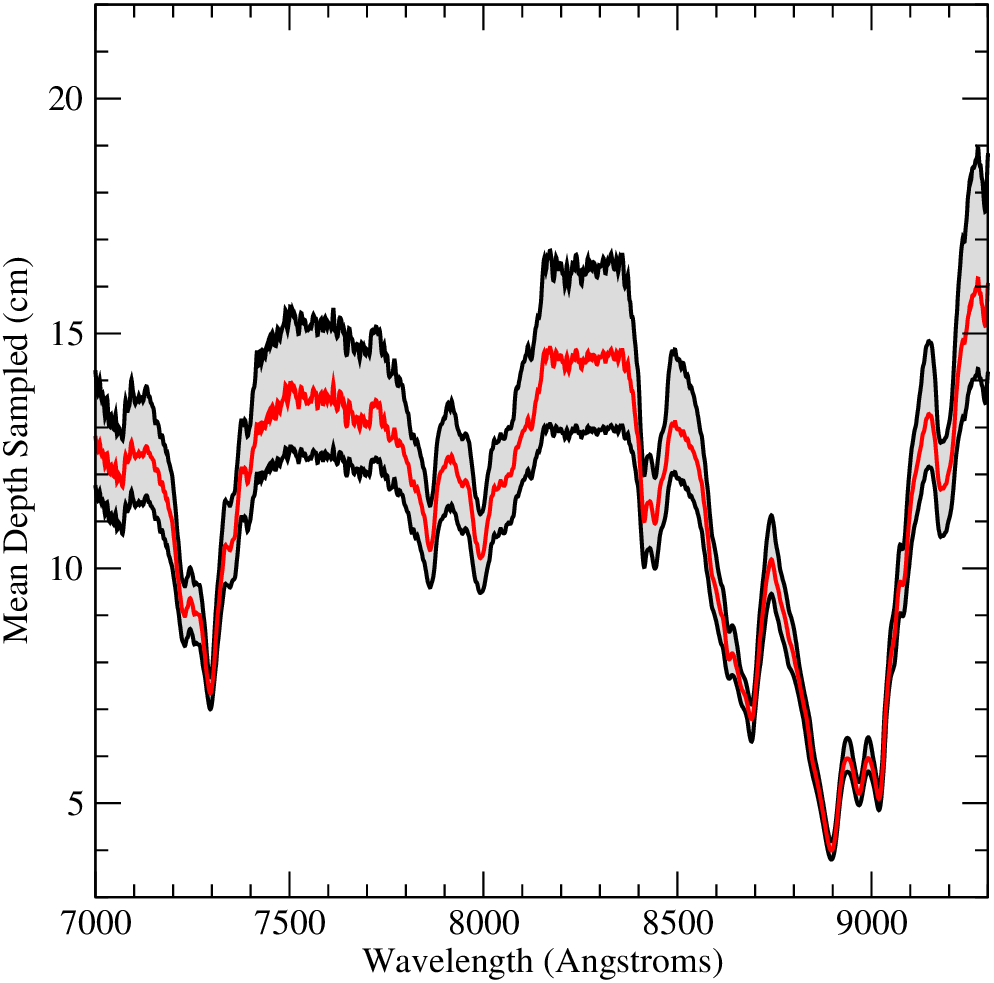}}
	\end{figure}

\end{document}